\documentstyle[aps]{revtex}
\begin{document}
\title{Electron impact excitation cross sections for hydrogen-like ions}
\author{Vladimir Fisher, Yuri V. Ralchenko, Vladimir Bernshtam, Alexander Goldgirsh,
and Yitzhak Maron}
\address{Faculty of Physics, Weizmann Institute of Science,\\
Rehovot 76100, Israel }
\author{Leonid A. Vainshtein}
\address{P. N. Lebedev Physical Institute, Russian Academy of Sciences, \\
Moscow 117924, Russia}
\author{Igor Bray}
\address{Electronic Structure of Materials Centre, School of Physical Sciences,\\
The Flinders University of South Australia, G.P.O. Box 2100,\\
Adelaide 5001, Australia}
\author{Helen Golten}
\address{Churchill College, Cambridge, United Kingdom{\it \ }}
\date{Received 20\ June, 1996}
\maketitle

\begin{abstract}
We present cross sections for electron-impact-induced transitions $%
n\rightarrow n^{\prime }$ in hydrogen-like ions $C^{5+}$, $Ne^{9+}$, $%
Al^{12+}$, and $Ar^{17+}$. The cross sections are computed by Coulomb-Born
with exchange and normalization (CBE) method for all transitions with $%
n<n^{\prime }<7$ and by convergent close-coupling (CCC) method for
transitions with $n<n^{\prime }<5$ in $C^{5+}$ and $Al^{12+}$. Cross
sections $1s\rightarrow 2s$ and $1s\rightarrow 2p$ are presented as well.
The CCC and CBE cross sections agree to better than 10\% with each other and
with earlier close-coupling results (available for transition $1\rightarrow
2 $ only). Analytical expression for $n\rightarrow n^{\prime }$ cross
sections and semiempirical formulae are discussed.

PACS number(s): 34.80.Kw
\end{abstract}

\section{INTRODUCTION}

Progress in many plasma physics fields (such as kinetics of incompletely
ionized plasmas, radiative hydrodynamics of non-equilibrium plasmas,
interpretation of spectroscopic measurements, etc.) depends on availability
and accuracy of electron-ion inelastic collision cross sections. In this
paper we present high-accuracy cross sections for electron-impact-induced
transitions between $n$-states of some hydrogen-like ions, namely, for
transitions $n\rightarrow n^{\prime }$ with $n<n^{\prime }<7$ in ions $%
C^{5+} $, $Ne^{9+}$, $Al^{12+}$, and $Ar^{17+}$. The cross sections are
denoted by $\sigma _{z,n,n^{\prime }}(x)$ where $z$ is the ion nuclear
charge, $n$ and $n^{\prime }$ are the principal quantum numbers of initial
and final states respectively ($n^{\prime }>n$), and

\[
x=\varepsilon /E_{nn^{\prime }} 
\]
is the ratio of the incident electron kinetic energy $\varepsilon \;$to
transition energy $E_{n,n^{\prime }}$. The cross sections are computed by
the Coulomb-Born with exchange and normalization (CBE) method \cite{1,2} and
by the convergent close-coupling (CCC) method \cite{3,4,5}.

In Sec. II, we present and compare the CCC and CBE cross sections together
with earlier close-coupling (CC) results \cite{6,7,8}. However, the CC cross
sections are published for transition $1\rightarrow 2$ only. We do not
include comparisons with broad-energy-range cross sections calculated by
less accurate methods (for assessment of theoretical data sources see papers
by Pradhan and Gallagher \cite{9} and Callaway\cite{10}). For use in
applications, Sec. III contains a table of coefficients which provide good
analytical fit to all $n\rightarrow n^{\prime }$ cross sections discussed in
Sec. II. Two semiempirical formulae for the cross section estimates are
discussed in Sections IV and V.

\section{THE CROSS\ SECTIONS}

The CBE cross sections were calculated by computer code ATOM \cite{1}.{\em \ 
}In this code, the exchange is accounted by method of orthogonalized
functions and the normalization is done by K-matrix method using one own
channel \cite{2}. Being installed on a personal computer, code ATOM provides
dozens of cross sections per day.

The CCC method is presented in Refs. \cite{3,4}. The basic idea of the CCC
approach to electron-atom and electron-ion collisions is to solve the
coupled equations arising upon expansion of the total wave function in a
truncated Laguerre basis of size $N$. This basis size is increased until
convergence to a desired accuracy is observed. The usage of the Laguerre
basis ensures that all states in the expansion are square-integrable, and so
gives a discretization of the target continuum as well as a good
representation of the target true discrete spectrum. For a sufficiently
large $N$ pseudoresonances, associated with the target continuum
discretization, diminish substantially so that no averaging is necessary.
The presented CCC calculations at all given energies are likely to be within
10\% of the true non-relativistic model solution for the considered
scattering systems. In general, the CCC method demonstrates excellent
agreement with experimental results available for various targets \cite
{3,4,5}.

The CCC and CBE computer codes produce cross sections for $nl\rightarrow
n^{\prime }l^{\prime }$ transitions. These cross sections are denoted below
by $\sigma _{z,nl,n^{\prime }l^{\prime }}(x)$. To obtain the cross sections
for the $n\rightarrow n^{\prime }$ transitions,$\ $partial cross sections $%
\sigma _{z,nl,n^{\prime }l^{\prime }}(x)$ are summed over $l^{\prime }$ and
averaged over $l$: 
\begin{equation}
\sigma _{z,n,n^{\prime }}(x)=\sum_{l=0}^{n-1}g_{nl}g_n^{-1}\sum_{l^{\prime
}=0}^{n^{\prime }-1}\sigma _{z,nl,n^{\prime }l^{\prime }}(x)\text{ }
\label{1}
\end{equation}
Here $g_{nl}$ and $g_n$ $=\sum_lg_{nl}$ are statistical weights of states $%
nl $ and $n$ respectively. Statistical averaging over $l$ (which is
reflected by the factor $g_{nl}g_n^{-1}$) is caused by uncertainty of $l$ in
initial quantum state given by $n$ only. Summation over $l^{\prime }$
ensures inclusion of all possible final states for a given $n^{\prime }$ of
interest.

The cross sections $\sigma _{z,n,n^{\prime }}(x)$ are presented in Figs. $%
1-8 $. One can see: (i) the CBE cross sections for all transitions with $%
n<n^{\prime }<7$ in ions $C^{5+}$, $Ne^{9+}$, $Al^{12+}$, and $Ar^{17+}$;
(ii) the CCC cross sections for all transitions with $n<n^{\prime }<5$ in
ions $C^{5+}$ and $Al^{12+}$; and (iii) the CC results available for
transition $1\rightarrow 2$ in ions $C^{5+}$, $O^{7+}$, $Ar^{17+}$ and $%
Fe^{25+}$ \cite{6,7,8}. To emphasize the scaling of the cross sections over $%
z^4$, we present scaled cross sections $z^4\sigma _{z,n,n^{\prime }}(x)$.
For $x\gg 1$ these are independent of $z$ to an accuracy better than 1\% .
For $x\sim 1$ there is a weak dependence on $z$ which results in small
deviations from the scaling. These deviations are to within $5\%$ for $%
z=10\div 18$ and to within $10\%$ for $z=6$ (except for transition $%
5\rightarrow 6$ in $C^{5+}$ which deviates by 17\% at $x\rightarrow 1$). In
present paper we limit our consideration by $z>5$ but it should be mentioned
that for $z\leq 5$ deviations from the $z^4$-scaling are more significant,
for example: $40\%$ for transition $1\rightarrow 2$ in $He^{+}$ and a factor
of 3 for transition $5\rightarrow 6$ in $He^{+}$.

To see the difference between CCC, CC, and CBE results with no influence of $%
z^4-$scaling, one may compare the cross sections related to the same ion.
However, the data available enables to compare the three approximations for
a single transition only, namely, for transition $1\rightarrow 2$ in $C^{5+}$
(Fig. 1). For this transition the CCC, CC, and CBE cross sections deviate
from each other by less than $8\%$. Generally, the CCC and CBE cross
sections of any transition studied deviate from each other by less than $%
10\%.$ Note, that in present work we do not compare broad--energy-range
cross sections to cross sections computed exclusively for narrow energy
intervals where contributions of resonances are significant (see, e.g.,
Refs. \cite{9,12,13}). Such fine detail is not necessary for the purpose of
most applications.

Figs. 9 and 10 demonstrate excellent agreement between the CCC, CC, and CBE
results for both monopole ($l^{\prime }=l$) and dipole ($l^{\prime }=l\pm 1$%
) channels contributing to transition $1\rightarrow 2$. At $x\approx 1$,
cross section $1s\rightarrow 2p$ is larger than $1s\rightarrow 2s$ by a
factor of 4; with $x$ the cross section $1s\rightarrow 2p$ goes down slower
than $1s\rightarrow 2s$ (namely, as $x^{-1}$ln$x$ vs. $x^{-1}$), therefore
the $1\rightarrow 2$ cross section is almost exclusively due to the dipole
channel. In general, in double sums (1) contributions from dipole
transitions dominate over contributions from non-dipole ones. Therefore,
properties of non-dipole cross sections are practically indistinguishable in
total cross sections $\sigma _{z,n,n^{\prime }}(x)$. Properties of partial
cross sections $\sigma _{z,nl,n^{\prime }l^{\prime }}(x)$ will be discussed
in separate paper \cite{14} which, in particular, demonstrates (i) strong
effect of electron exchange for energy range $x<3$, and (ii) an increase of
this effect with multipole order $\mid l^{\prime }-l\mid $.

\section{FITTING FORMULA}

To simplify a use of the cross sections in applications, we fitted them by
analytical function. Taking account for the $z^4-$scaling and analysis
presented in Ref. \cite{15} the fitting function is chosen to be

\begin{equation}
\sigma _{z,n,n^{\prime }}^f(x)=\pi a_0^2z^{-4}x^{-1}(\alpha _{n,n^{\prime
}}\ln x+\beta _{n,n^{\prime }}x^{-1}\ln x+\gamma _{n,n^{\prime }}+\delta
_{n,n^{\prime }}x^{-1}+\zeta _{n,n^{\prime }}x^{-2})\text{ }.  \label{2}
\end{equation}
Here $a_0$ is the Bohr radius. Coefficients $\alpha _{n,n^{\prime }},\beta
_{n,n^{\prime }},\gamma _{n,n^{\prime }},\delta _{n,n^{\prime }},$and $\zeta
_{n,n^{\prime }}$ are listed in Table 1. The fitting function $\sigma
_{z,n,n^{\prime }}^f(x)$ provides an accuracy to better than 10\% for any $x$
for all transitions studied ( $z=6\div 26$) except for transition $%
5\rightarrow 6$ in $C^{5+}$. For this transition, the fit is less accurate
because at $x\rightarrow 1$ the cross section deviates by $17\%$ from total
scaling over $z^4$ (Fig. 8).

\section{THE\ VAN\ REGEMORTER\ FORMULA}

Expression (2) may be used in applications which require high accuracy of
the cross sections. For estimates, it is desirable to have a simpler
expression which does not use a table of coefficients. Frequently, such
estimates are based on the Van Regemorter formula \cite
{2,9,11,16,17,18,19,20}. For transitions between $n$-states this formula may
be presented as follows

\begin{equation}
\sigma _{z,n,n^{\prime }\;}^{VR}(x)=\pi a_0^2\frac{8\pi f_{nn^{\prime }}}{%
\sqrt{3}}\frac{{\cal R}^2}{E_{nn^{\prime }}^2}\;\frac{G(x)}x\;.\;  \label{3}
\end{equation}
Here $f_{nn^{\prime }}\;$is the absorption oscillator strength, ${\cal R}$ $%
=13.6$ eV is the Rydberg energy unit, and $G(x)\;$is the effective Gaunt
factor which may be treated as a fitting function of order unity.

To find an accurate expression for the $n$-independent function $G(x),$ we
first use fitting function (2) and Eq. (3) to introduce the Gaunt factor $%
G_{nn^{\prime }}(x)$ for each of transitions studied: 
\[
G_{nn^{\prime }}(x)=x\sigma _{z,n,n^{\prime }}^f(x)\left( \pi a_0^2\frac{%
8\pi f_{nn^{\prime }}}{\sqrt{3}}\frac{{\cal R}^2}{E_{nn^{\prime }}^2}\right)
^{-1}.
\]
These Gaunt factors for all transitions with $n<n^{\prime }<7$ are shown by
dotted curves in Fig. 11. The curves are not labeled because they are shown
only for demonstrating the small spread of functions $G_{nn^{\prime }}(x)$
near their mean function 
\begin{equation}
G(x)=0.349\ln x+0.0988+0.455x^{-1}  \label{4}
\end{equation}
which is shown by bold solid curve. We recommend function (4) as effective
Gaunt factor for the Van Regemorter formula (3).

Bold dot-dash curves in Fig. 11 show $\pm $50\% corridor around this $G(x).$
One can see that for $x\approx 1$ some dotted curves deviate from $G(x)$ by
more than a factor of 2, but for $x>5$ a spread of dotted curves is smaller
(to within $\pm $50\%), and for $x=100\div 1000$ the spread is to within
20\%.

\section{THE VSY FORMULA}

The Van Regemorter formula (3) is most accurate for $x\gg 1.$ Therefore,
this formula fits applications which require accurate account of
suprathermal electrons (e.g., pulsed power devices, subpicosecond lasers,
solar flares). However, there are non-Maxwellian plasmas with overpopulated
low-energy part of the electron distribution function, e.g., plasmas
produced by UHF devices or by lasers with non-relativistic intensity of
radiation (for our example, it is enough to have a free electron oscillation
energy less than mean energy of the electron chaotic motion). Estimates of
kinetic coefficients for such plasmas require cross sections accurate in
low-energy range ($x=1\div 10$). Semiempirical formula suitable for this
case was suggested by Vainshtein, Sobel'man, and Yukov \cite{11,2}. We
rewrite it as follows 
\begin{equation}
\sigma _{z,n,n^{\prime }\;}^{VSY}(x)=\frac{\pi a_0^2}{\sqrt{n^{\prime }}}%
\left( \frac{{\cal R}}{I_n-I_{n^{\prime }}}\right) ^2\left( \frac{%
I_{n^{\prime }}}{I_n}\right) ^{3/2}\frac{F(x)}x\;\;  \label{5}
\end{equation}
where $I_n$ is the optical electron binding energy and 
\begin{equation}
F(x)=14.5\ln x+4.15+9.15x^{-1}+11.9x^{-2}-5.16x^{-3}  \label{6}
\end{equation}
is the fitting function which provides good fit to CC, CCC, and CBE cross
sections discussed above. Being interested in an easy-to-use formula, we
looked for function $F(x)$ independent on $z$, $n$, and $n^{\prime }$
although initially \cite{11,2} functions $F(x)$ were fitted to each of the
transition studied. Using the expression $I_n={\cal R}z^2n^{-2}$ for the
binding energy, formula (5) may be presented as follows

\begin{equation}
\sigma _{z,n,n^{\prime }\;}^{VSY}(x)=\frac{\pi a_0^2}{z^4}n^7\sqrt{n^{\prime
}}\left[ (n^{\prime })^2-n^2\right] ^{-2}\frac{F(x)}x.  \label{7}
\end{equation}

Function (6) is shown by bold solid curve in Fig. 12. Dotted curves
demonstrate functions $F_{nn^{\prime }}(x)$ obtained by replacing $\sigma
_{z,n,n^{\prime }\;}^{VSY}(x)$ in expression (7) by functions (2):

\[
F_{nn^{\prime }}(x)=\frac{z^4x\sigma _{z,n,n^{\prime }\;}^f(x)}{\pi a_0^2}%
\frac{\left[ (n^{\prime })^2-n^2\right] ^2}{n^7\sqrt{n^{\prime }}}. 
\]
These dotted curves are not labeled because they are shown only for
demonstrating their small deviation from function $F(x)$. Bold dash curves
show $\pm $30\% corridor around $F(x)$ while bold dot-dash curves show $\pm $%
50\% corridor. One can see that for $x<2$ the VSY formula is accurate to
within 30\%. For $x=1\div 10$ this formula is accurate to within 50\%. For
larger electron energy ($x>10$), estimates of the cross sections are more
accurate if based on the Van Regemorter formula (3).

\section{CONCLUSIONS}

We have presented CCC and CBE cross sections for electron-impact-induced
transitions $n\rightarrow n^{\prime }$ in hydrogen-like ions $C^{5+}$, $%
Ne^{9+}$, $Al^{12+}$, and $Ar^{17+}$ ($n<n^{\prime }\leq 6$).

With coefficients given in Table 1, expression (2) fits all CC, CCC, and CBE
cross sections available for $z=6\div 26$ to better than 10\%, except for
transition $5\rightarrow 6$ in $C^{5+}$ which deviates from total $z^4$%
-scaling by 17\%.

In general, a scaling of the cross sections over $z^4$ is accurate to within
a few percent for ions with large nuclear charge ($z\gg 1)$ but for ions
with $z\sim 1,$ a deviation from the scaling is significant (at $x\sim 1$).
For $x\gg 1$ the $z^4$-scaling is accurate for all ions and transitions. For
each $z,$ the accuracy of the scaling is higher for larger transition energy.

Semiempirical formulae (3) and (7) together provide an accuracy to within
50\% for any energy: for $x<2$ the VSY formula (7) is accurate to within
30\%; for $2<x<10$ this formula is accurate to within 50\%; for $x>10$ an
accuracy to better than 50\% is provided by the Van Regemorter formula (3).

\section{ACKNOWLEDGMENTS}

This work was supported by the Israel Academy of Science, The Ministry of
Science and the Arts, and the Ministry of Absorption. In part (L.V.) the
work was supported by Moscow International Science and Technology Center
(grant 07-95) and Russian Basic Research Fond (grant 94-2-05371). Research
of I. Bray is sponsored in part by the Phillips Laboratory, Air Force
Materiel Command, USAF, under cooperative agreement number F29601-93-2-0001.
The views and conclusions contained in this document are those of the
authors and should not be interpreted as necessarily the official policies
or endorsements, either expressed or implied, of Phillips Laboratory or the
U.S. Government.

\ \ \

\section{Figure Captions}

{\bf Figure 1a. }Scaled cross sections for transition $1\rightarrow 2$ in
hydrogen-like ions with $z=6\div 26$.

{\bf Figure 1b. }Fragment of Figure 1a.

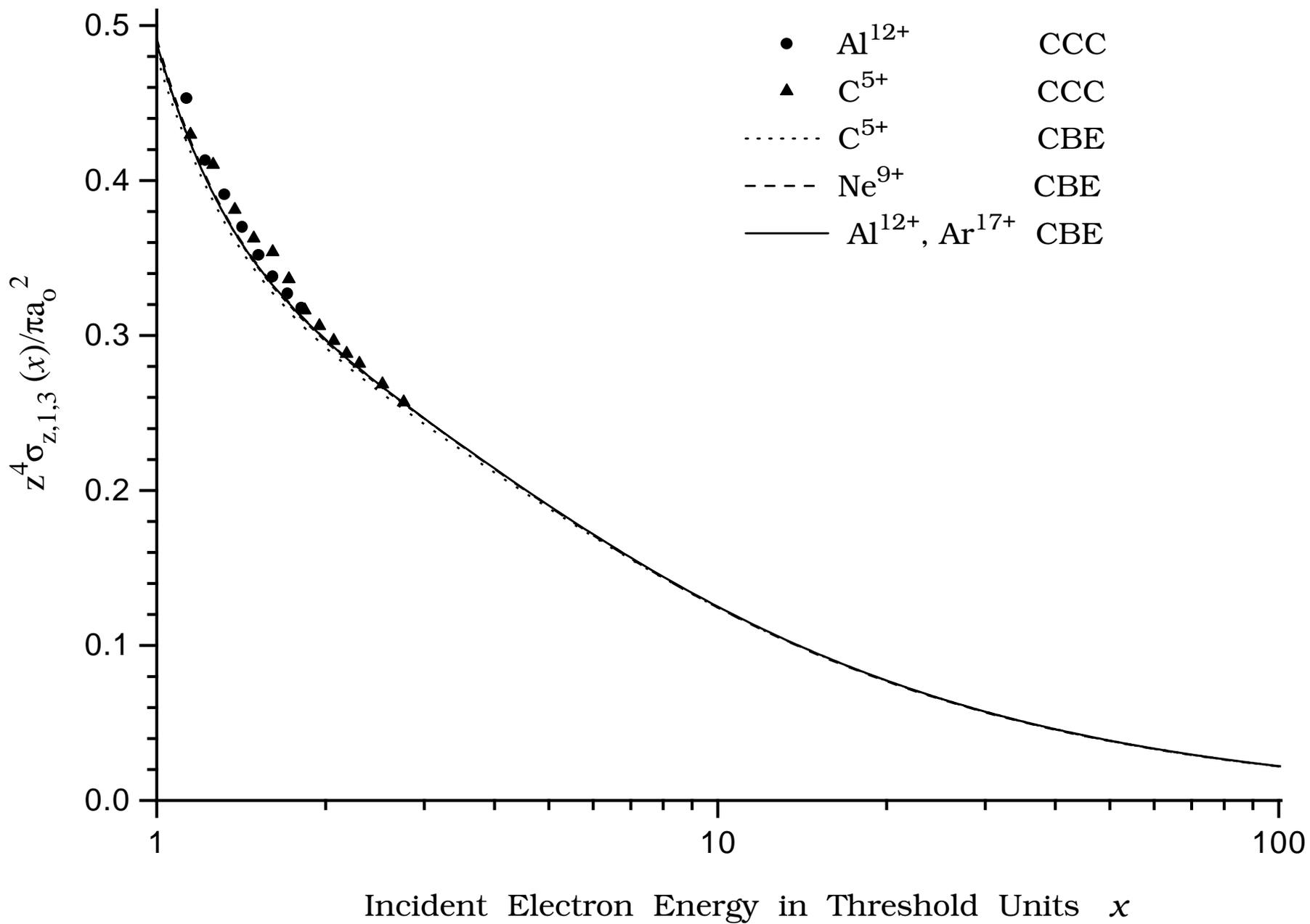
{\bf Figure 2. }Scaled cross sections for transition $1\rightarrow 3$ in
hydrogen-like ions with $z=6\div 18$.

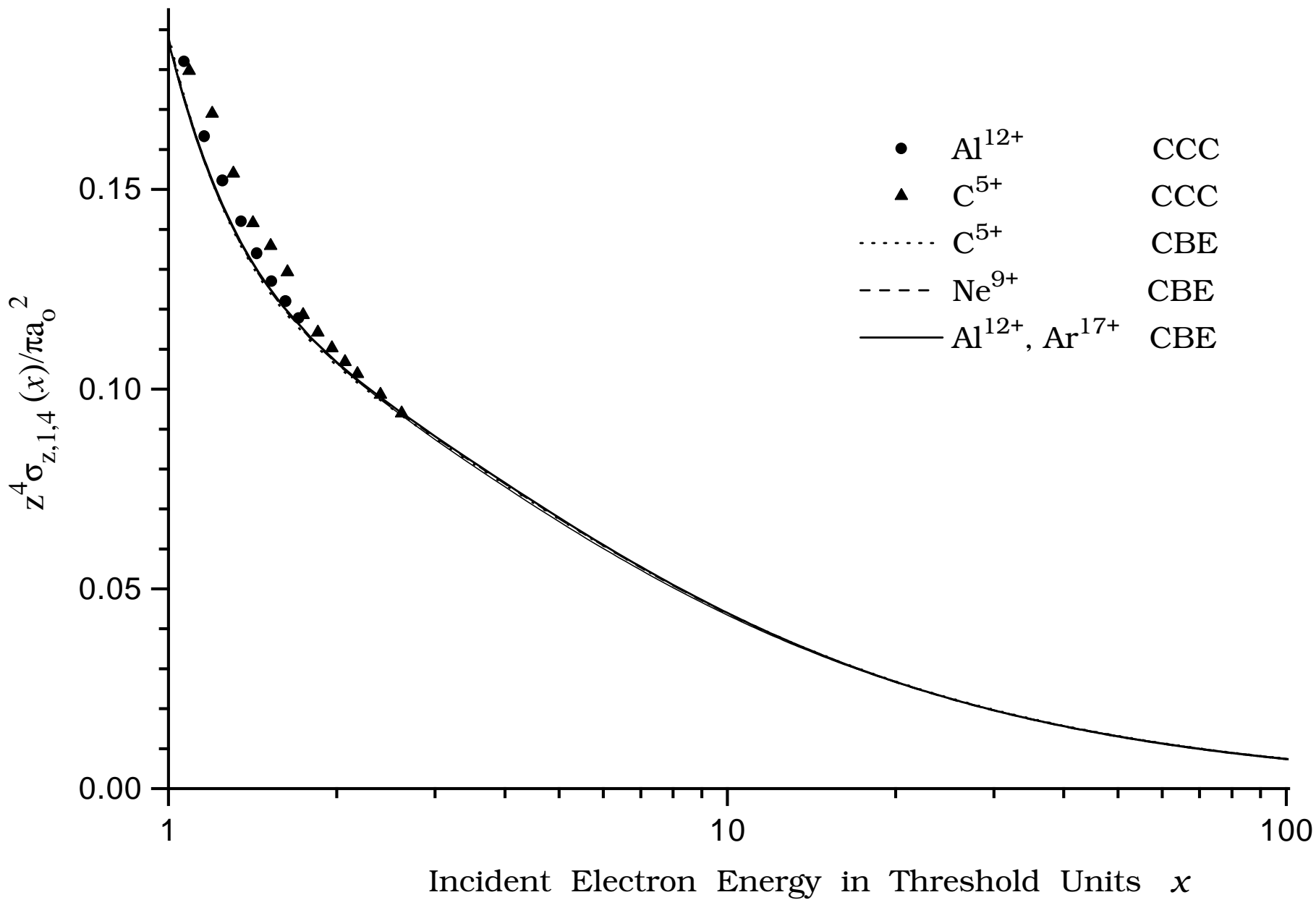
{\bf Figure 3. }Scaled cross sections for transition $1\rightarrow 4$ in
hydrogen-like ions with $z=6\div 18$.

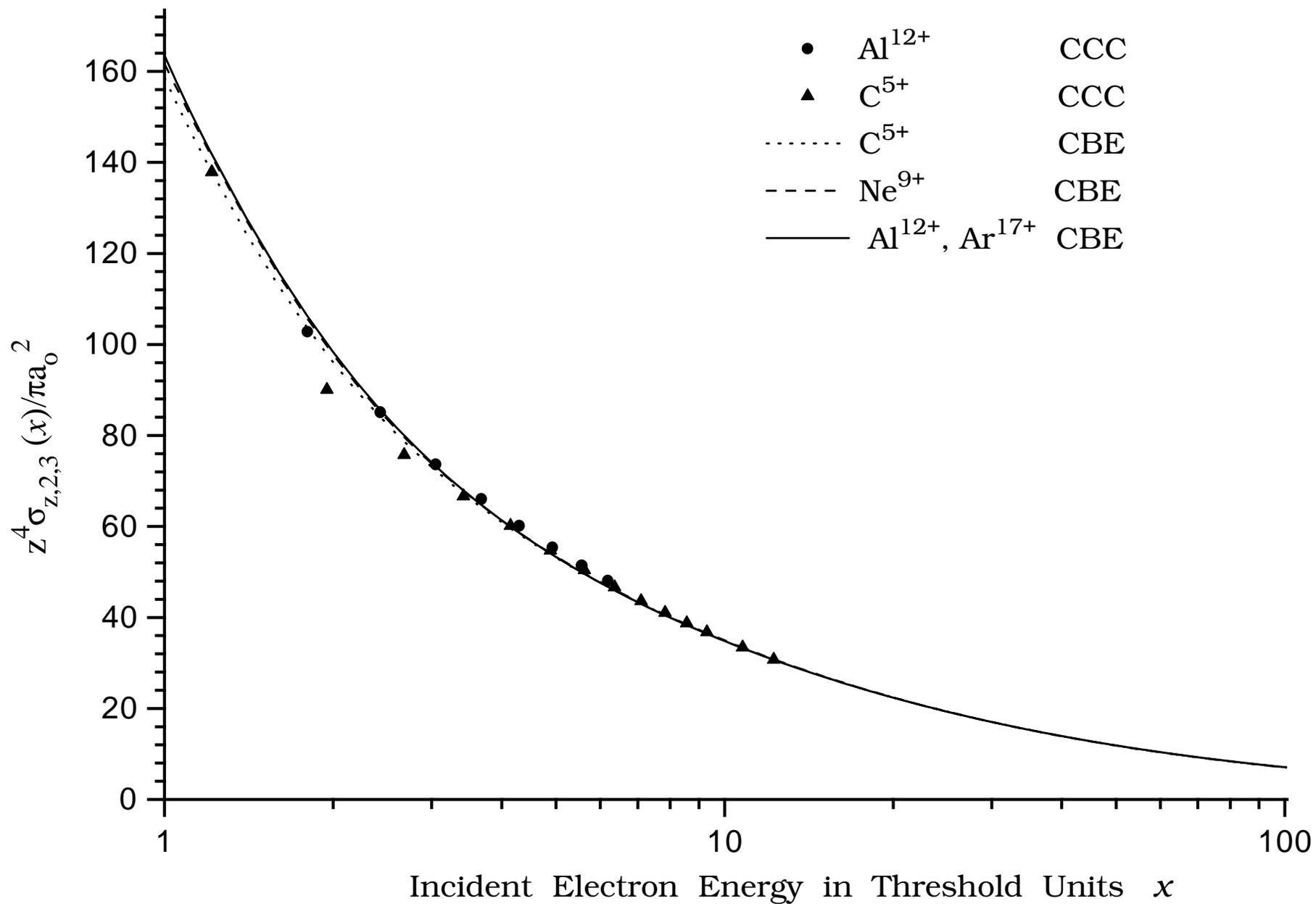
{\bf Figure 4. }Scaled cross sections for transition $2\rightarrow 3$ in
hydrogen-like ions with $z=6\div 18$.

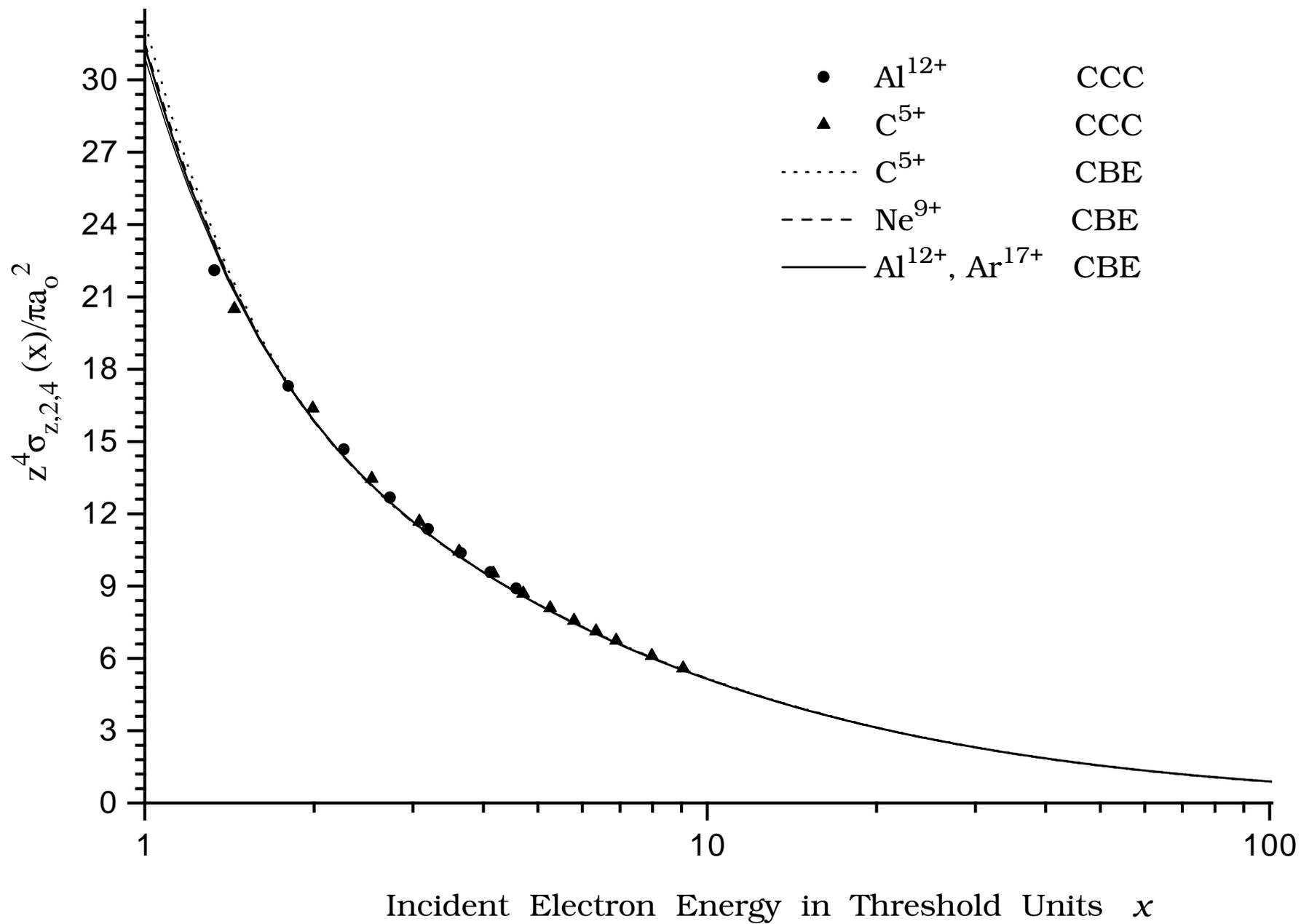
{\bf Figure 5. }Scaled cross sections for transition $2\rightarrow 4$ in
hydrogen-like ions with $z=6\div 18$.

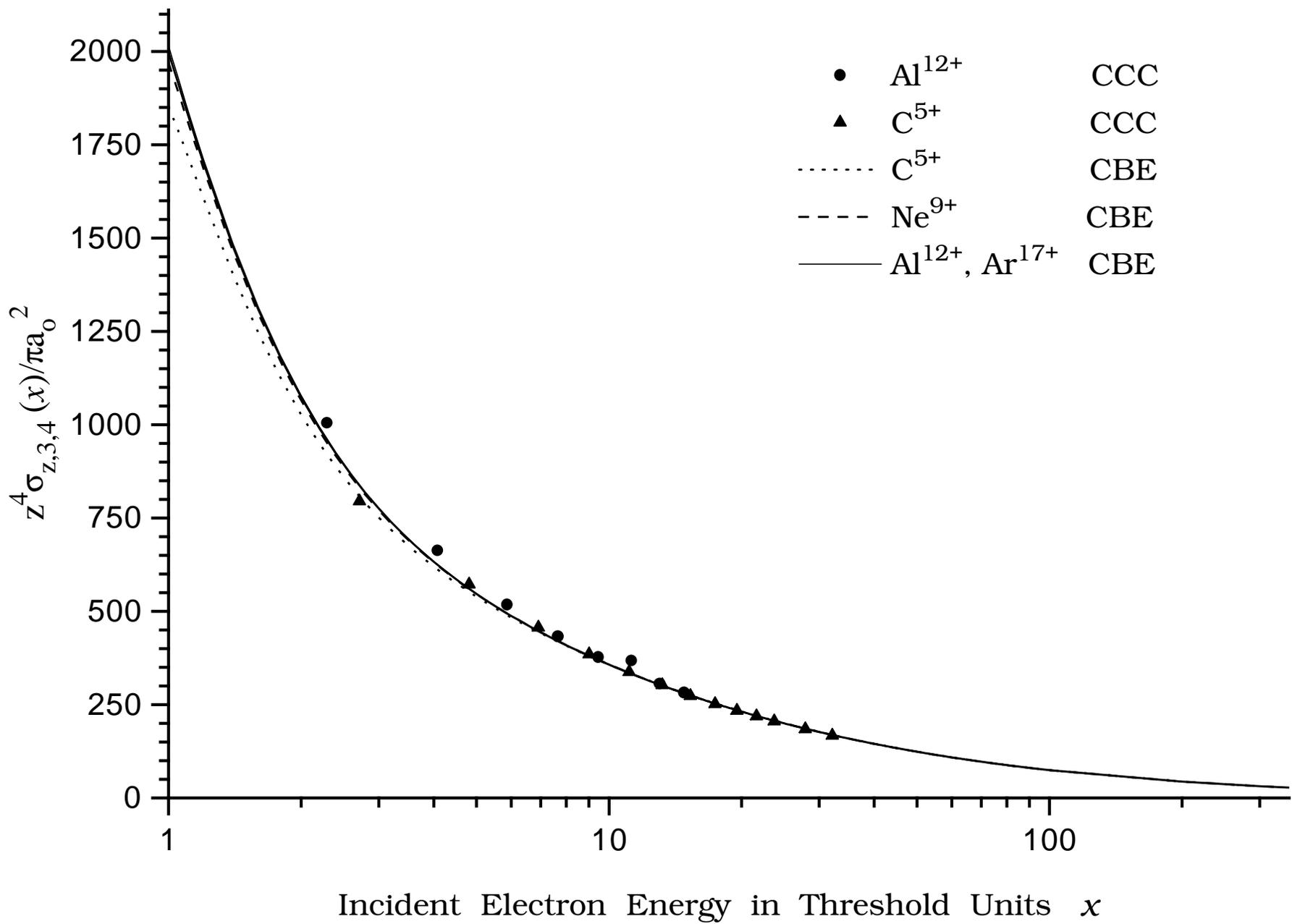
{\bf Figure 6. }Scaled cross sections for transition $3\rightarrow 4$ in
hydrogen-like ions with $z=6\div 18$.

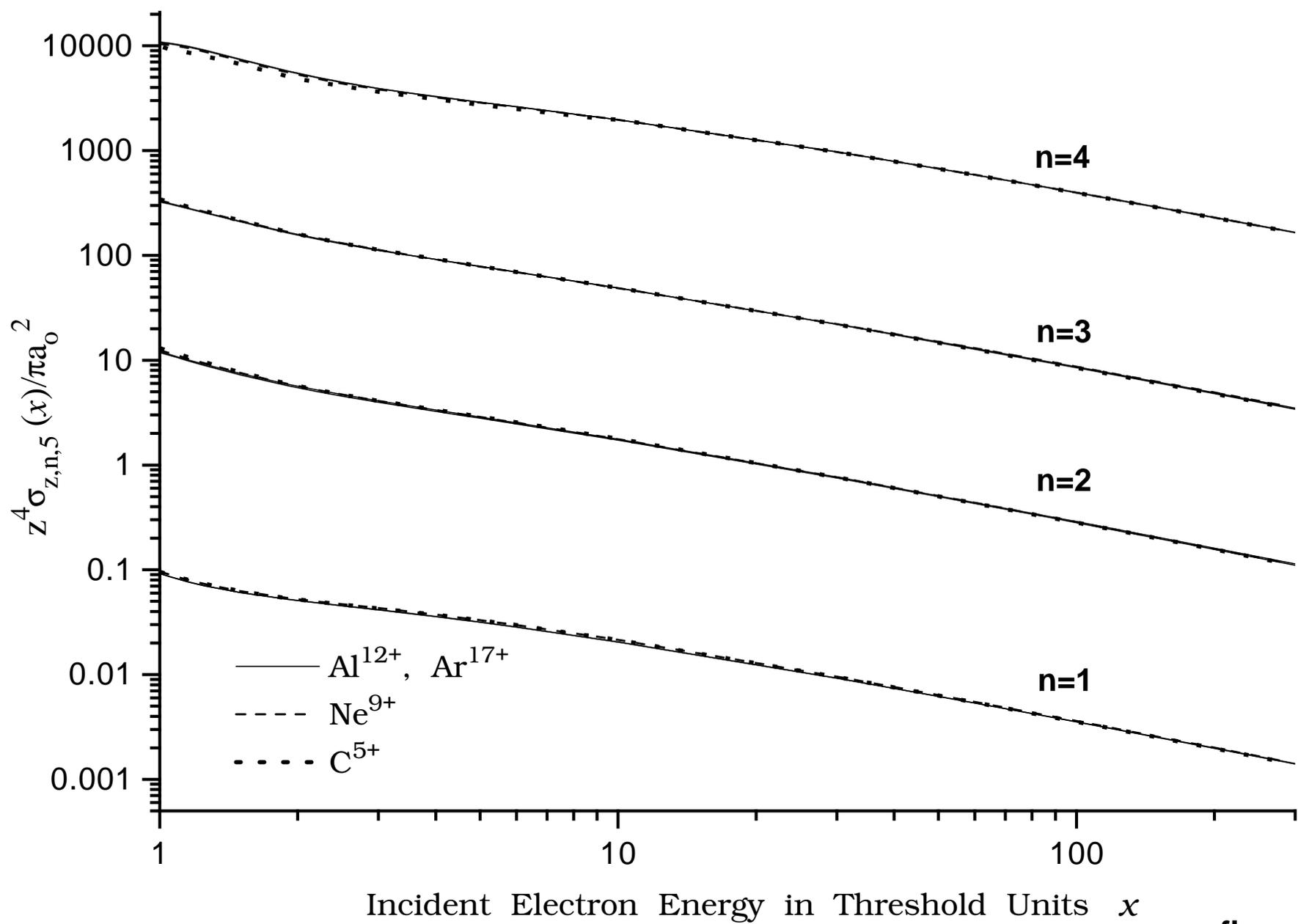
{\bf Figure 7. }Scaled CBE cross sections for transitions $n\rightarrow 5$
in hydrogen-like ions with $z=6\div 18.$

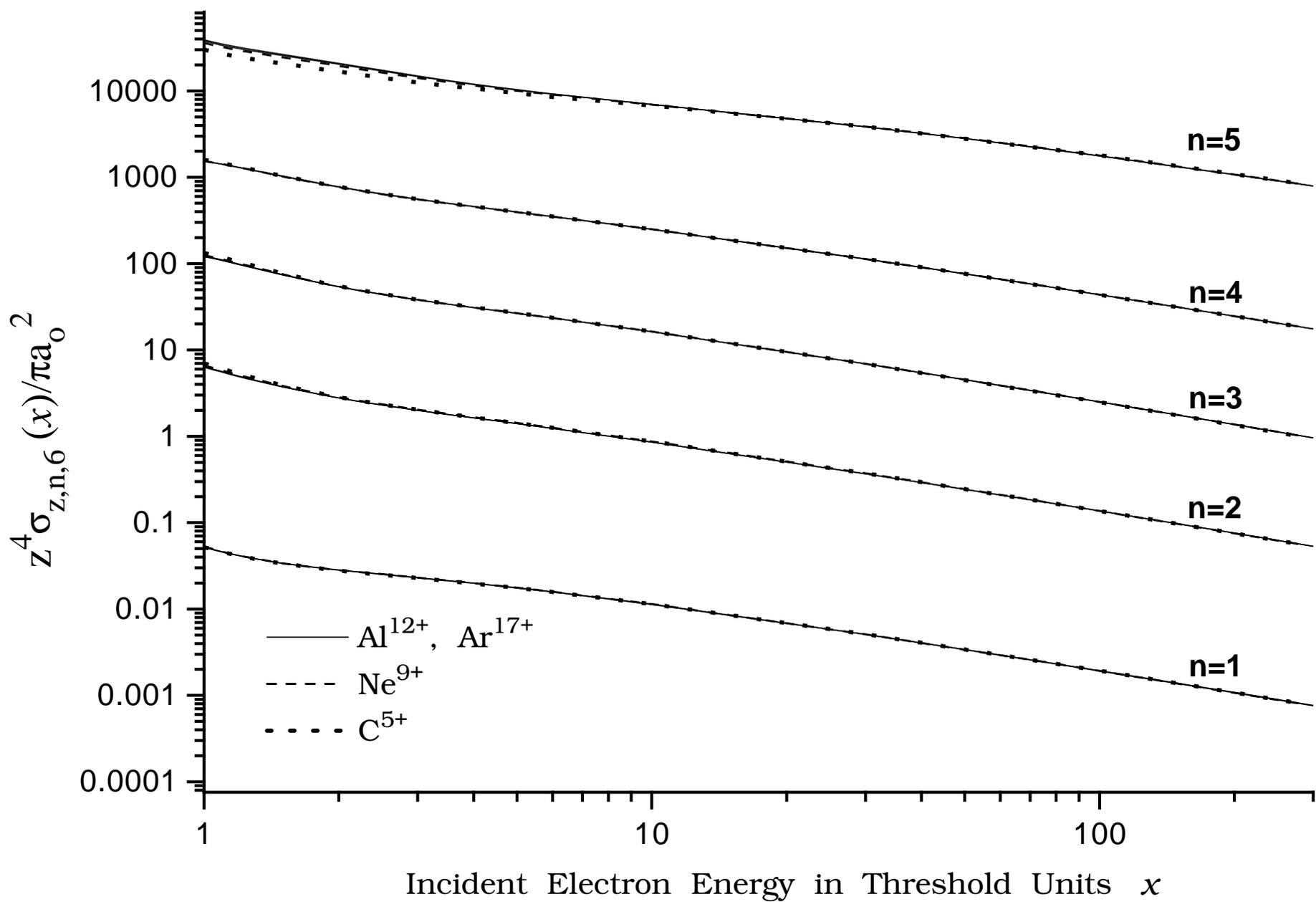
{\bf Figure 8. }Scaled CBE cross sections for transitions $n\rightarrow 6$
in hydrogen-like ions with $z=6\div 18.$

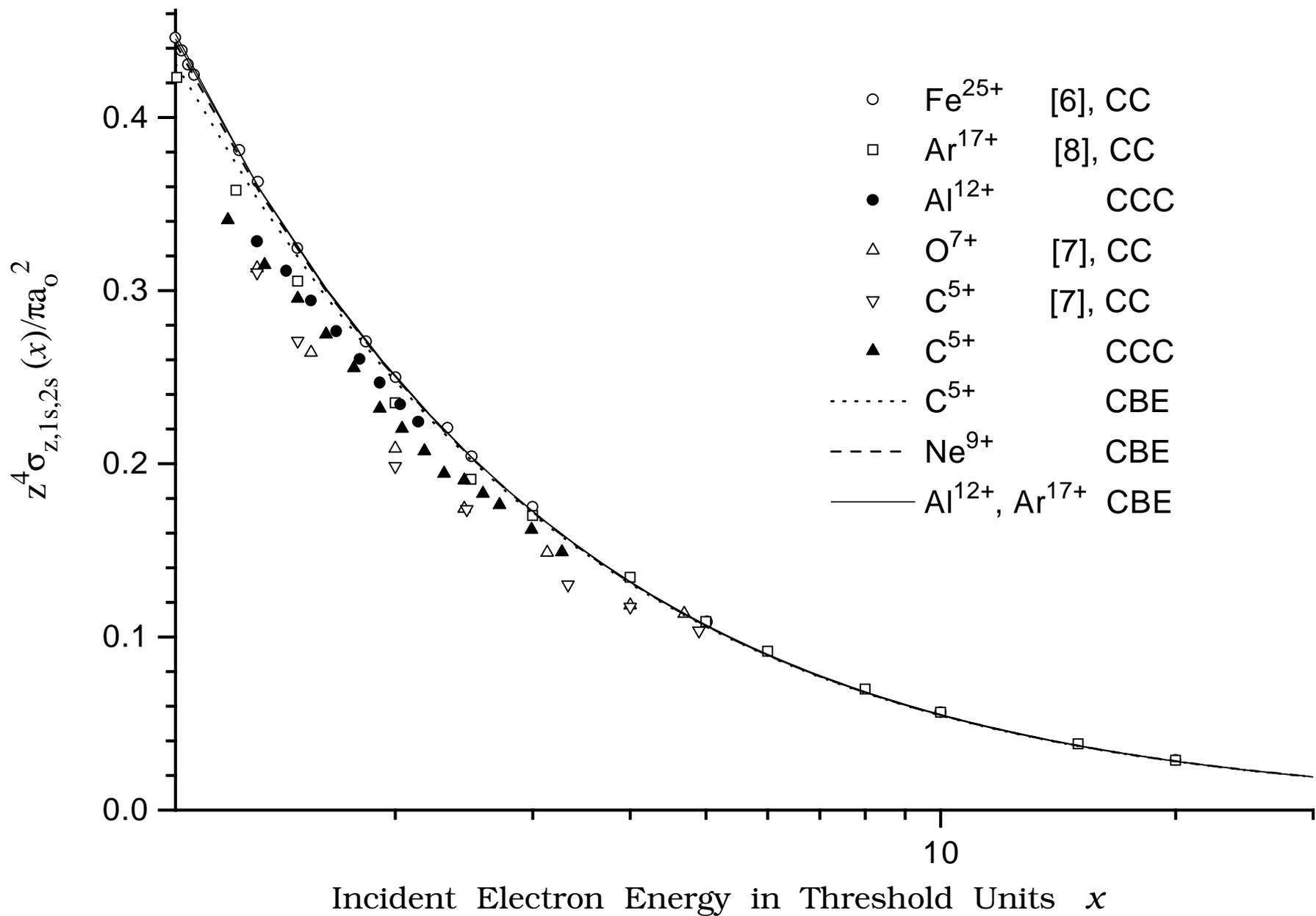
{\bf Figure 9. } Scaled CC, CCC, and CBE cross sections for transition $%
1s\rightarrow 2s$ in hydrogen-like ions with $z=6\div 26$.

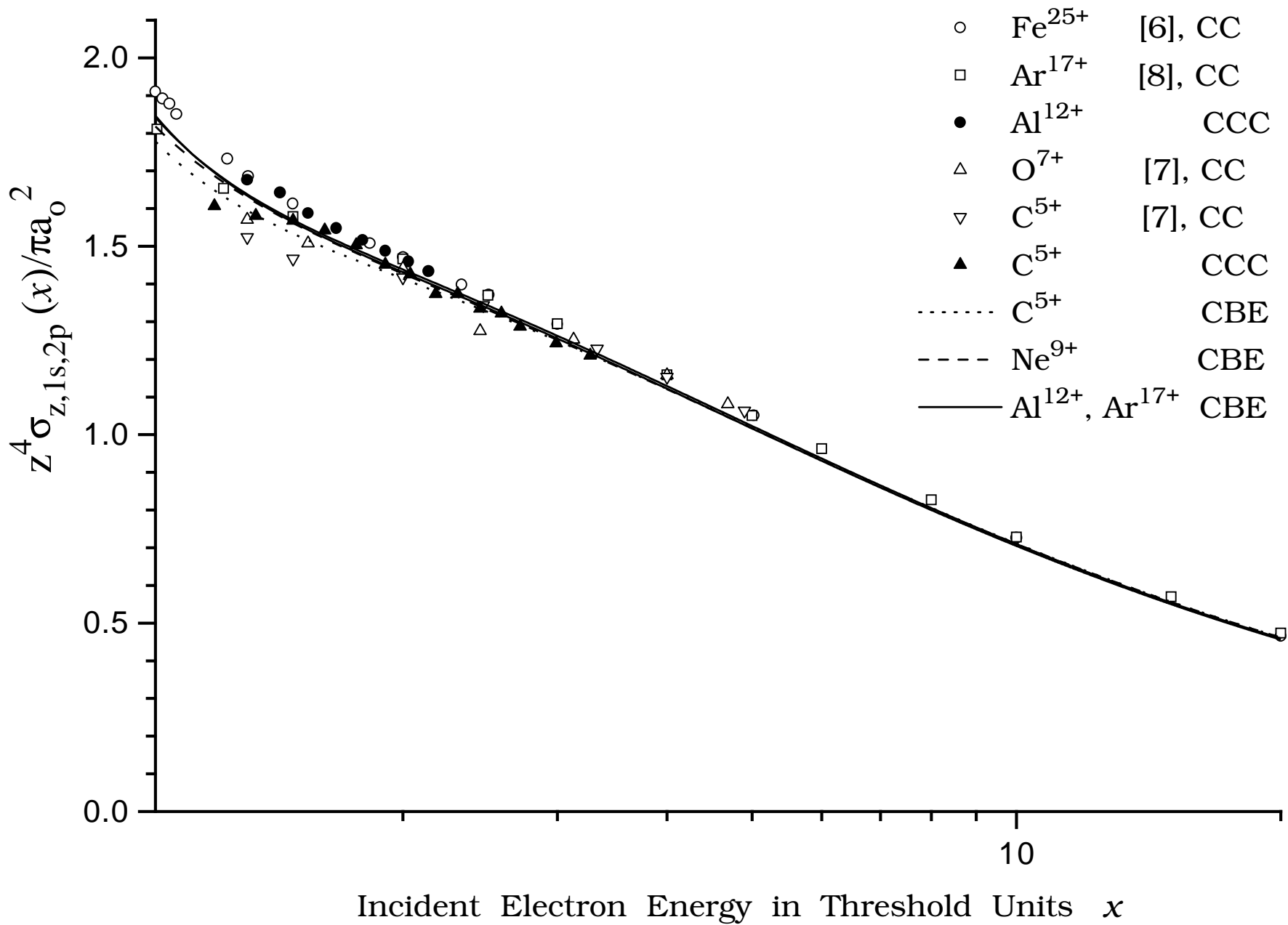
{\bf Figure 10. }Scaled CC, CCC, and CBE cross sections for transition $%
1s\rightarrow 2p$ in hydrogen-like ions with $z=6\div 26$.

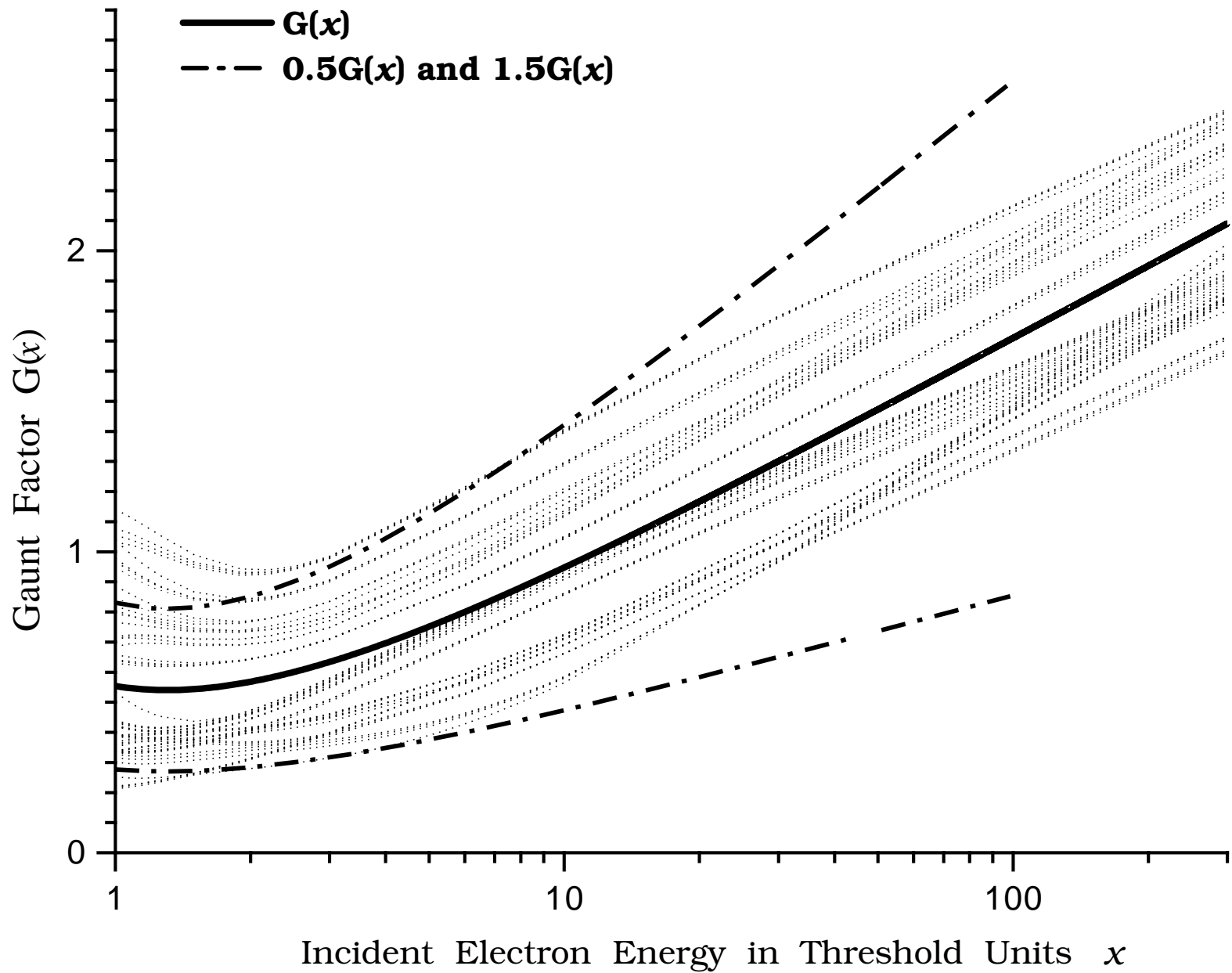
{\bf Figure 11. }Gaunt{\bf \ }factor $G(x)$ for the Van Regemorter formula
(3). Dotted curves demonstrate functions $G_{nn^{\prime }}(x).$

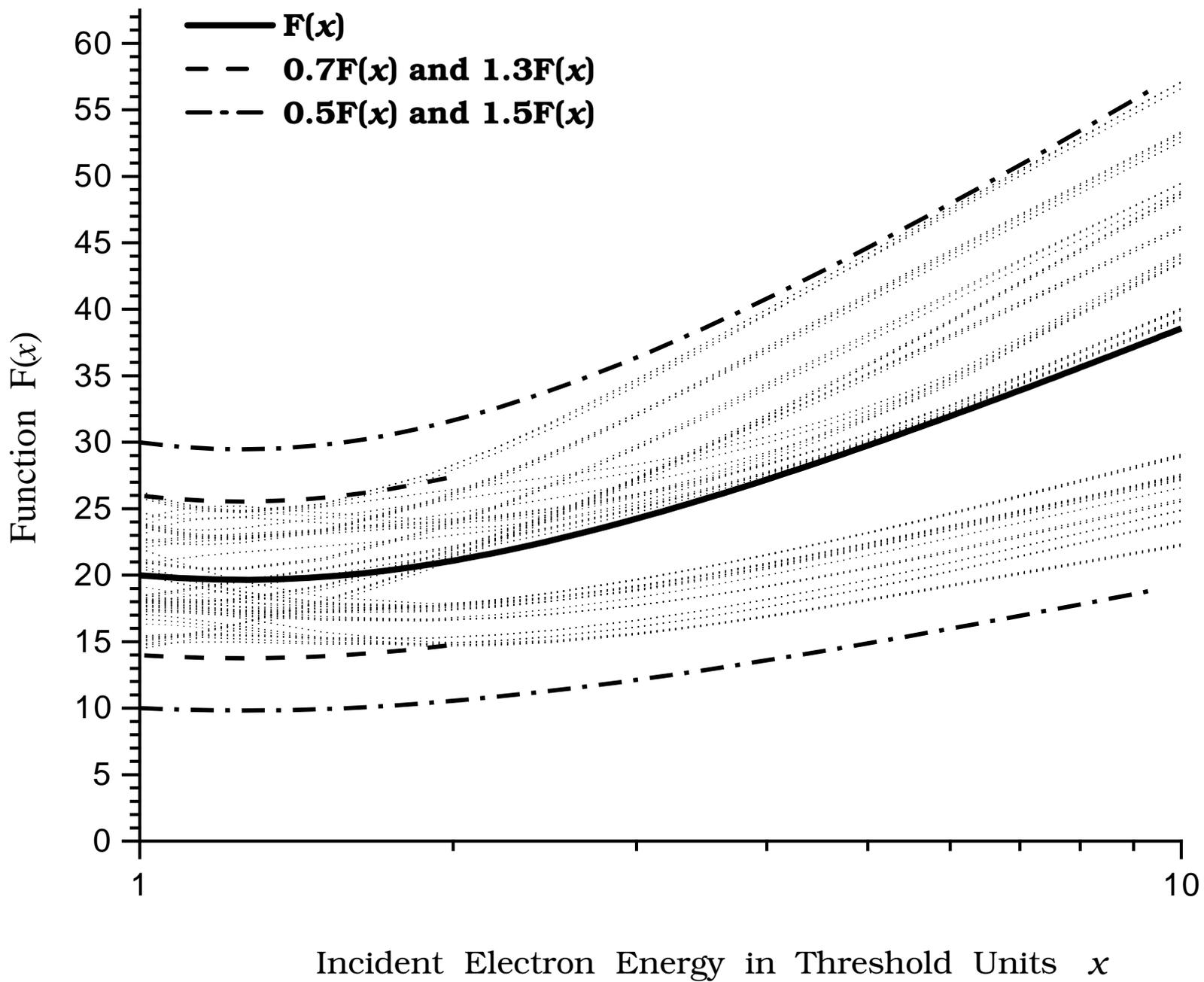
{\bf Figure 12. }Function $F(x)$ for the VSY formula (7). Dotted curves
demonstrate functions $F_{nn^{\prime }}(x).$

\section{Table 1}

\begin{tabular}{ccccccc}
\hline\hline
n & n$^{\prime }$ & $\alpha _{nn^{\prime }}$ & $\beta _{nn^{\prime }}$ & $%
\gamma _{nn^{\prime }\text{ }}$ & $\delta _{nn^{\prime }}$ & $\zeta
_{nn^{\prime }}$ \\ \hline\hline
1 & 2 & 3.22 & 0.357 & 0.00157 & 1.59 & 0.764 \\ 
1 & 3 & 0.452 & 0.723 & 0.0291 & -0.380 & 0.834 \\ 
1 & 4 & 0.128 & -0.0300 & 0.163 & -0.150 & 0.185 \\ 
1 & 5 & 0.0588 & -0.0195 & 0.0803 & -0.0649 & 0.0776 \\ 
1 & 6 & 0.0321 & -0.0115 & 0.0458 & -0.0374 & 0.0441 \\ 
2 & 3 & 173 & 46.7 & -94.5 & 358 & -102 \\ 
2 & 4 & 16.7 & -8.32 & 12.6 & 22.5 & -3.44 \\ 
2 & 5 & 4.47 & -5.54 & 8.10 & 3.52 & 0.820 \\ 
2 & 6 & 1.94 & -2.69 & 4.43 & 0.461 & 1.54 \\ 
3 & 4 & 1880 & 204 & -1280 & 4860 & -1620 \\ 
3 & 5 & 164 & -236 & 128 & 458 & -247 \\ 
3 & 6 & -2.71 & -468 & 268 & 161 & -304 \\ 
4 & 5 & 4800 & -55100 & 18100 & 41400 & -48800 \\ 
4 & 6 & 456 & -4530 & 2230 & 2870 & -3570 \\ 
5 & 6 & 75000 & 247000 & -172000 & 84700 & 123000 \\ \hline
\end{tabular}

{\bf Table 1.} Coefficients $\alpha _{nn^{\prime }}$, $\beta _{nn^{\prime }}$%
, $\gamma _{nn^{\prime }\text{ }}$, $\delta _{nn^{\prime }}$, and$\ \zeta
_{nn^{\prime }}$ for fitting function (2).

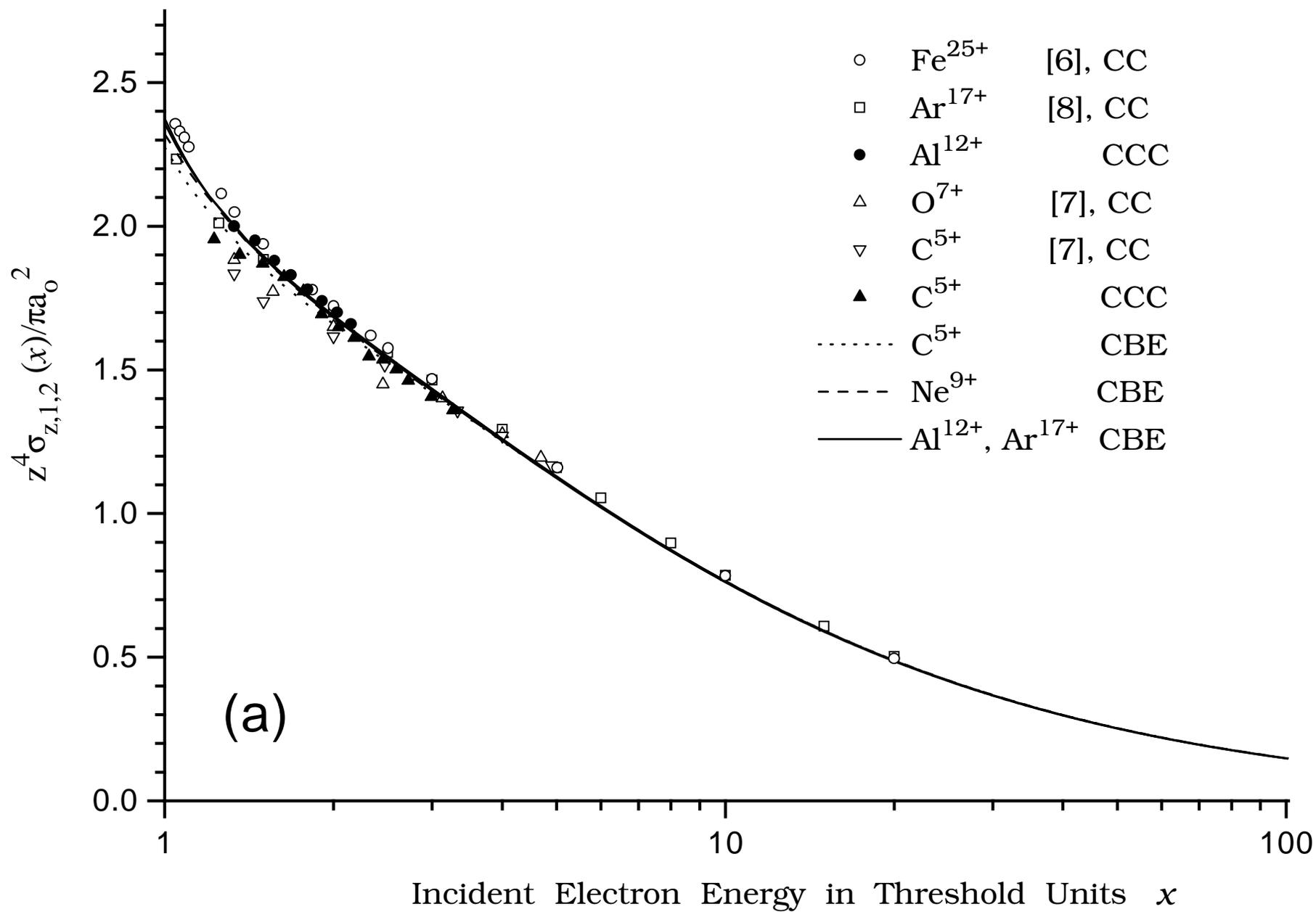
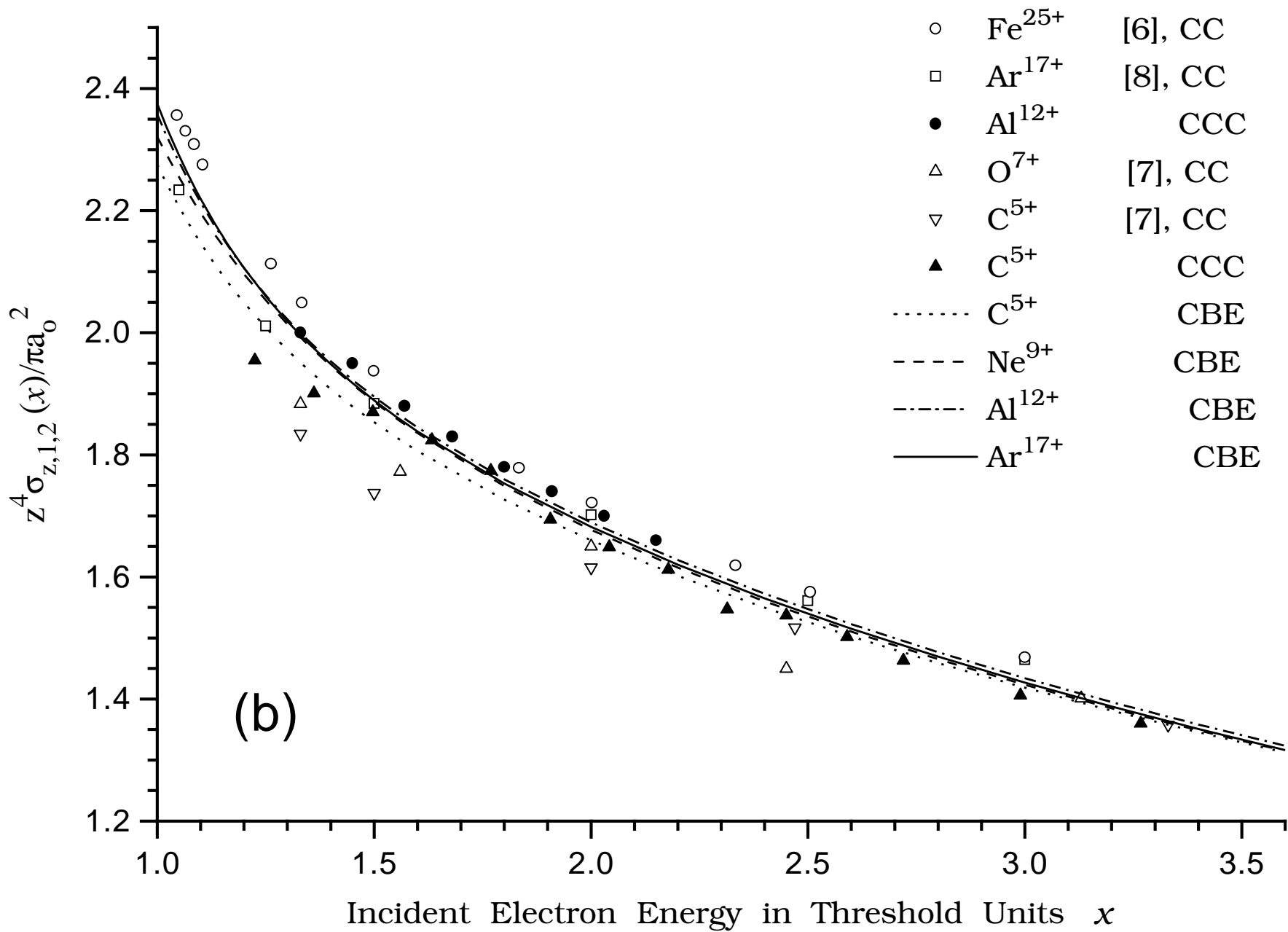
\end{document}